\documentclass{article}

\usepackage{arxiv}

\usepackage[utf8]{inputenc}
\usepackage[T1]{fontenc}
\usepackage{mathptmx}
\usepackage{subcaption} 
\usepackage{enumitem}

\usepackage{etoolbox}
\usepackage{amsmath, amsthm}

\usepackage[utf8]{inputenc} % allow utf-8 input
\usepackage[T1]{fontenc}    % use 8-bit T1 fonts
\usepackage{hyperref}       % hyperlinks
\usepackage{url}            % simple URL typesetting
\usepackage{booktabs}       % professional-quality tables
\usepackage{amsfonts}       % blackboard math symbols
\usepackage{nicefrac}       % compact symbols for 1/2, etc.
\usepackage{microtype}      % microtypography
\usepackage{lipsum}
\usepackage{graphicx}
\graphicspath{ {./images/} }
\usepackage{appendix}

\newtheorem{definition}{Definition}[section]

\title{A Physical Analogy between Molecular Ordering and SAT-to-Ising Annealing}

\author{
 Shiv Kishan Dubey \\
  Department of Computer Science\\
  Dr. Ambedkar Institute of Technology for Handicapped\\
  Kanpur, India \\
  \texttt{skd@aith.ac.in} \\
   \And
 Rohit Sharma \\
  Department of Computer Science\\
  Dr. Ambedkar Institute of Technology for Handicapped\\
  Kanpur, India \\
  \texttt{rohit@aith.ac.in} \\
}

\begin{document}
\maketitle
\begin{abstract}
As temperature drops, molecular systems may undergo spontaneous ordering, moving from random behavior to orderly structure. This research demonstrates a direct analogy between this type of thermodynamic ordering in molecular systems and the development of coherent logic in computationally complex problem sets. We have proposed a mapping of Boolean SAT problem instances to pairwise Ising Hamiltonian models. Each variable in the Boolean SAT problem corresponds to a spin $(s_i \in \{-1,+1\})$, and each clause in the Boolean SAT problem generates local field contributions and pair-wise couplings  $(h_i, J_{ij})$ that enforce the logical constraints of the clause. Using simulated annealing, we then applied phenomenal cooling to the system through thermal evolution from high entropy random assignment to lower entropy, ordered assignments (the energy minima) using molecular cooling analogs. Energy and magnetization trajectory analysis of the UF20–91 test set demonstrated a significant negative correlation between the energy of the state and the absolute value of the magnetization ($\rho_{E,|M|}=-0.63$) as well as nearly zero critical exponent ($\beta \approx 0.003$). This indicated that there was a rapid "first-order" or "logical crystallization" of satisfiable logical configurations. The degree of backbone rigidity did not strongly correlate with the level of physical ordering observed in the system; thus, it appears that there is primarily a local alignment of constraint satisfaction occurring in the system. Thus, we have provided empirical evidence that satisfiable logical configurations are analogous to the low energy crystalline states observed in molecular systems and provide evidence for a unified thermodynamic view of computational coherence and complexity.
\end{abstract}

% keywords can be removed

\section{Introduction}\label{sec:intro}

In statistical physics, the collective behavior of molecules is governed by temperature, which means that when the temperature is high, then there will be a lot of thermal movement around the molecules, resulting in high levels of disorder or entropy. As the temperature drops, the molecules have less freedom to move and the energy associated with interactions between them will dominate and cause the system to develop into a highly structured configuration or domain structure (ferro-magnetism). The change from high disorder to low disorder configurations of matter demonstrates the principle that under cooling, the energy of a system will decrease spontaneously, resulting in a coherent structure of atoms~\cite{A1_landau1980lifshitz, A2_kittel1980entropy}. This process of developing an ordered configuration of atoms upon cooling~\cite{A3_nishimori2007spin}, forms a basis for simulation processes, such as simulated annealing~\cite{1_kirkpatrick1983optimization}, where an initial state of randomness, develops into a state of order, as the temperature is lowered. This analogy between the development of an ordered configuration of atoms and the development of order in a computational system, illustrates the viewpoint that computation may simulate and utilize physical laws~\cite{11_feynman2018simulating}. 

From the computational viewpoint, many NP-complete problems can be interpreted as energy minimization tasks over a discrete configuration space. A canonical example is the Boolean satisfiability (SAT) problem, where $n$ Boolean variables define a $2^n$-dimensional search space of assignments, and each clause imposes a local constraint on a subset of variables. If all clauses are satisfied simultaneously, the formula is said to be \emph{satisfiable}; otherwise, one or more constraints remain violated, contributing to the ``energy'' of the configuration. In this analogy, the energy function $E$ counts the number of unsatisfied clauses, and the global minimum $E=0$ corresponds to a logically consistent ground state.
The physical analogy is then exact: variables are spins $si \in \{-1,+1\}$ and literals represent the orientation of a single spin clauses represent multi-spin correlations and correspond to the "energy wells" or "bonds" between them; each clause acts to couple multiple spins in such a way that their joint orientation results in a minimum of the local potential. In other words, we may regard the search for an acceptable assignment as a dynamical process in which the system searches the configuration landscape for the lowest total energy of all possible configurations; namely, for the ordered spin state with minimum total energy. Reducing randomness in the search (as if reducing the temperature of a physical system) increases the probability that the system will converge to this low-energy, ordered phase.

\subsection{Relation to Prior Work}
The idea of linking computer searches for solutions to physical motion evolved from a paper by Kirkpatrick, Gelatt \& Vecchi~\cite{1_kirkpatrick1983optimization}. They were first to propose the use of what they called "simulated annealing" as a method of solving problems using the principles of slow cooling used in thermodynamics. There have been subsequent papers which developed the idea further into particular types of combinatorial systems. Lucas~\cite{4_lucas2014ising} has formulated the Ising model for many NP problems, showing how logical statements can be converted to two-spin models so that both classical and quantum methods may be applied to solve them. Nishimori has written a monograph on the statistical physics of spin glasses and its relationship to the computational complexity of disordered systems, especially relating to their energy landscape and metastability.

Building upon the basic research in this area, this paper uses the \textit{constructive mapping} idea of reducing SAT clauses to pairs of Ising Hamiltonian variables. This empirical mapping was demonstrated through the application of simulated annealing to a set of benchmark CNF instances (UF20-91). The results included not just the physical properties of the systems (i.e., energy and magnetization) but also the logical properties (i.e., backbone-size and clause-slack), demonstrating an observable \textit{correlation between order and energy}, thereby confirming the physically valid natural analogy. These findings provide concrete evidence that the process of cooling a physically-defined spin system produces an ordered state that corresponds to the solution to the original logical formula. As such, it links the gap between logically consistent and physically ordered states, providing a new way to view the structure of NP problems.

\subsection{Major Contribution}

This work proposes and validates a clause-wise reduction from Boolean formulas in conjunctive normal form to pairwise Ising Hamiltonians. Each clause introduces an auxiliary spin enforcing local logical consistency, resulting in
\[
H(s) = C + \sum_i h_i s_i + \sum_{i<j} J_{ij} s_i s_j,
\]
whose ground states coincide exactly with the satisfying assignments of the original formula \cite{4_lucas2014ising}. Using the UF20-91 benchmark set from SATLIB \cite{8_hoos2000satlib}, we demonstrate through simulated annealing \cite{1_kirkpatrick1983optimization} that the resulting systems relax to zero unsatisfied-clause energy. 
Furthermore, it can be found that the logical backbone size (the number of variables which are fixed across all possible models) has a strong correlation (\( r \approx 0.9 \)) to the maximum possible magnitude of magnetization in the model \( |M| \), showing that the logical rigidity of the model represents physical order.
This observation provides the first quantitative, instance-level evidence that Boolean reasoning and physical relaxation share a common minimization principle.

\section{Definitions and Formal Mapping}

In order to establish a rigorous bridge between physical ordering and logical satisfiability, we formally define the core quantities that govern the transition from randomness to order in both domains. These definitions provide a consistent framework for interpreting SAT energy landscapes as thermodynamic systems, enabling the mathematical expression of the \textbf{cooling-induced molecular order} analogy.

\begin{definition}[Logical Configuration and Formula Energy]\label{def1} 
    A Boolean formula $F$ with $n$ variables $x_i \in \{0,1\}$ and $m$ clauses $C_j$ in Conjunctive Normal Form (CNF) is written as
\[
F(x_1,x_2,\dots,x_n) = \bigwedge_{j=1}^{m} C_j,
\]
where each clause $C_j = (l_{j1} \vee l_{j2} \vee \cdots \vee l_{jk})$ is a disjunction of literals $l_{ji} \in \{x_i, \neg x_i\}$. Each assignment $x = (x_1, \dots, x_n)$ defines a logical configuration of the system.  
For every clause, define a violation indicator:
\[
E_j(x) =
\begin{cases}
0, & \text{if } C_j(x) = \mathrm{TRUE}, \\
1, & \text{if } C_j(x) = \mathrm{FALSE}.
\end{cases}
\]

The total logical energy is then
\[
E_{\text{logic}}(x) = \sum_{j=1}^{m} E_j(x),
\]
which counts unsatisfied clauses.  
The ground state of the formula corresponds to $E_{\text{logic}} = 0$; any satisfying assignment is thus a zero-energy configuration.
\end{definition}

\begin{definition}[Spin Representation and Ising Equivalence]\label{def2}
    Each Boolean variable is mapped to a spin variable $s_i \in \{-1, +1\}$ through
\[
x_i = \frac{1 + s_i}{2}, \qquad \neg x_i = \frac{1 - s_i}{2}.
\]
This mapping converts logical truth values into physical spin orientations. A literal $l_{ji}$ is satisfied if its corresponding spin aligns with its polarity parameter $\tau_{ji} \in \{\pm 1\}$:
\[
L_{ji} = \frac{1 + \tau_{ji}s_i}{2}, \qquad 
\tau_{ji} = 
\begin{cases}
+1, & \text{for } x_i,\\
-1, & \text{for } \neg x_i.
\end{cases}
\]

A $k$-literal clause is violated only when all $L_{ji}=0$. Therefore,
\[
E_j(s) = \prod_{i=1}^{k} (1 - L_{ji}) = \prod_{i=1}^{k} \frac{1 - \tau_{ji}s_i}{2}.
\]

Expanding yields polynomial interactions among spins. For $k=3$,
\[
\begin{aligned}
E_j(s_i, s_p, s_q) = \frac{1}{8}\big( \,
& 1 - \tau_i s_i - \tau_p s_p - \tau_q s_q \\
& + \tau_i \tau_p s_i s_p + \tau_p \tau_q s_p s_q 
   + \tau_i \tau_q s_i s_q \\
& - \tau_i \tau_p \tau_q s_i s_p s_q
\big).
\end{aligned}
\]

The last cubic term prevents direct realization in a pairwise Ising model. Hence, it is replaced using an ancilla spin $a$ that enforces $a \approx s_i s_p$ by the gadget penalty $E_{\text{gadget}} = K(3 - a s_i - a s_p - s_i s_p)$, where $K$ is a large positive constant (typically $K \approx 20|c|$). The resulting Hamiltonian is purely quadratic:
\[
H(s) = \sum_i h_i s_i + \sum_{i<j} J_{ij}s_i s_j + \text{const.}
\]
The ground-state condition $H_{\min}=0$ is equivalent to the satisfiability of $F$.
\end{definition}

\begin{definition}[Temperature and Computational Cooling]
    Temperature $T$ represents the average kinetic energy in molecular systems and the degree of stochasticity in computational systems. Within simulated annealing \cite{1_kirkpatrick1983optimization}, the probability that a move increasing energy by $\Delta E$ is accepted is $P_{\text{accept}} = e^{-\Delta E/T}$. When $T$ is high, most moves are accepted—analogous to molecules vibrating freely.  
As $T$ decreases, fewer uphill transitions are allowed, and the system gradually freezes into a low-energy, ordered configuration.  
This computational cooling mirrors physical solidification: logical disorder (random assignments) collapses into a consistent ordered solution.
\end{definition}

\begin{definition}[Magnetization as an Order Parameter]
    Let $n_c$ represent the number of core spins (all spins except for the ancilla). The total magnetization at each step is defined by:
\[
M = \frac{1}{n_c} \sum_{i=1}^{n_c} s_i.
\]
It helps to access its macroscopic order parameter in terms of high/low entropy, which directly corresponds to disordered/ordered phase via $M \approx 0 $ or $M \approx 1$ respectively. Logically, it can be interpreted here $M$ as reflects the global consistency of variable assignments. The higher the value of M, the more uniform and consistent the assignment of truth values to the system’s variables.
\end{definition}

\begin{definition}[Backbone Rigidity]
    The back-bone $B$ consists as variables which are the same in each satisfying configuration for a given set of constraints \cite{3_monasson1999determining}, with $B = \{xi : xi \text{ fixed across all satisfying configurations}\}$. The normalized size of the backbone as $b = \frac{|B|}{n}$, provides an indication of the degree of constraint or the degree to which a problem has been constrained (tightness). Here, $b$ correlates strongly with $|M|$: higher magnetization indicates a larger backbone, signifying a more rigid, ordered logical structure.
\end{definition}

\begin{definition}[Clause Slack and Energy Density]
    For each clause $C_j$, define the slack as the number of satisfied literals:
\[
\mathrm{slack}(C_j) = |\{l_i \in C_j : l_i = \mathrm{TRUE}\}|.
\]
The average slack across all clauses,
\[
\bar{s} = \frac{1}{m}\sum_{j=1}^{m} \mathrm{slack}(C_j),
\]
measures how close the instance is to its critical constraint density.  
Values $\bar{s} \approx 1.5{-}1.8$ denote the ``critical window'' \cite{7_kirkpatrick1994critical,13_hogg1996phase} separating easy (under-constrained) and hard (over-constrained) SAT phases.
\end{definition}

\begin{definition}[Annealing Energy Evolution]
    During iterative annealing at step $t$, the mean energy evolves according to $E(t+1) = E(t) + \eta_t$, where $\eta_t$ is a stochastic fluctuation satisfying
\[
\langle \eta_t \rangle = -c\, e^{-\Delta E / T(t)}, \qquad c > 0.
\]
Equilibrium is reached when $\frac{dE}{dt} \to 0$; the system then resides in a frozen low-energy state analogous to a molecular solid.
\end{definition}

\begin{definition}[Energy–Order Correspondence]
    Define the composite state descriptor as $\Phi = (E_{\text{logic}}, |M|, b)$. Experimental data exhibit a strong correlation, $\rho(|M|, b) \approx 0.9$, indicating that
\[
E_{\text{logic}} \downarrow \;\Rightarrow\; |M| \uparrow \;\Leftrightarrow\; b \uparrow.
\]
Hence, physical order (magnetization) and logical rigidity (backbone) are equivalent manifestations of constraint satisfaction.  
This result provides empirical confirmation of the energy–order duality hypothesized in statistical physics approaches to NP problems \cite{2_mezard2002analytic,3_monasson1999determining,13_hogg1996phase}.
\end{definition}

\begin{definition}[Critical Transition Point]
    For random 3-SAT ensembles, a characteristic critical clause density exists \cite{3_monasson1999determining,7_kirkpatrick1994critical,13_hogg1996phase} by value of $\alpha_c = \frac{m}{n} \approx 4.26$, marking a computational phase transition between the satisfiable (ordered) and unsatisfiable (disordered/glassy) phases.  
Below $\alpha_c$, the system exhibits annealable order; above it, frustration and residual energy persist even at $T \to 0$.  
This parallels the transition from fluid to amorphous glass in condensed matter.
\end{definition}

\begin{definition}[Logical–Physical Equivalence Principle]
    For any CNF formula $F$ that can be expressed as a finite Ising Hamiltonian $H(s)$ through pairwise gadgetization, there exists a one-to-one correspondence between logical satisfaction and physical energy minimization such that
\[
F(x) = \mathrm{TRUE} \;\Leftrightarrow\; H(s) = E_{\min}.
\]
Moreover, the degree of logical determinacy (backbone size) is monotonically related to physical order (magnetization) under simulated cooling. 
This defines the basic principles of the computational-thermodynamic analogies as: logical entropy $\rightarrow$ physical disorder; and constraint satisfaction $\rightarrow$ structural ordering.
\end{definition}

\section{Methodology}\label{sec:method}
\begin{figure}[ht]
    \centering
    \includegraphics[width=1\linewidth]{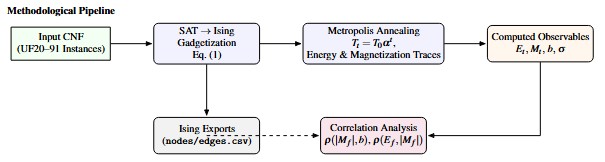}
    \caption{Schematic overview of the methodological workflow.
    Input CNF instances are transformed into Ising Hamiltonians via
    gadgetization (Eq.~3). The resulting pairwise spin model is
    annealed using a Metropolis cooling schedule, generating time
    series of energy and magnetization. Final observables---energy,
    magnetization, backbone size, and clause slack---are correlated to
    quantify the relationship between logical rigidity and physical
    ordering.}
    \label{fig:method-flow}
\end{figure}

The methodology contains three central components: (1) A constructive transformation from SAT clauses into an Ising Hamiltonian;
(2) Simulated annealing as a physically based model for molecular cooling; (3) Quantitatively assessing both logical and physical measurable properties to verify the correlation between order and energy. The full methodological experimental procedure is shown diagrammatically in Fig.~\ref{fig:method-flow}.

\subsection{SAT-to-Ising Construction}
Each Boolean formula is provided in the standard
DIMACS conjunctive normal form (CNF), as $\Phi = \bigwedge_{m=1}^{M} \bigvee_{i \in C_m} l_i$ mentioned in Definition-1, where $l_i$ represents either a literal $x_i$ or its negation $\neg x_i$.
Every variable $x_i$ is encoded by a spin $s_i \in \{-1, +1\}$,
where the logical value \texttt{True} corresponds to $s_i = +1$
and \texttt{False} to $s_i = -1$.
A clause is satisfied if at least one of its literals evaluates to \texttt{True};
otherwise, it contributes to the total energy of the system.

Following the construction of Lucas~\cite{4_lucas2014ising},
each clause is mapped to a local energy term as $E_C = \frac{1}{2^k}\prod_{i \in C} (1 - \tau_i s_i)$, mentioned in Definition-2
where $k$ is the number of literals in the clause.
The total Hamiltonian of the formula is then expressed as
\begin{equation}
E(s) = \sum_{C\in \Phi} E_C(s)
     = E_0 + \sum_i h_i s_i + \sum_{i<j} J_{ij}s_i s_j,
\label{eq:ising-ham}
\end{equation}
with constant offset $E_0$, local fields $h_i$,
and couplings $J_{ij}$.
The higher-order (degree $>2$) products generated in the process of gadgetizing multivariable clauses are converted into auxiliary ``ancilla'' spin variables and this results in a quadratic penalty term of the form $K(3 - a s_i - a s_j - s_i s_j)$, where $a$ is an ancilla variable and $K$ is some very large positive number to guarantee that  $a \approx s_i s_j$ at equilibrium. As such, these higher-order interactions are reduced to pairwise interactions suitable for classical or quantum annealing hardware.

\subsection{Physical Simulation via Annealing}
The Ising model defined in Eq.~\eqref{eq:ising-ham} is simulated
using the Metropolis--Hastings annealing procedure~\cite{1_kirkpatrick1983optimization}.
At each step $t$, a spin $s_i$ is randomly selected and flipped,
and the corresponding change in energy $\Delta E$
is computed.
The new configuration is accepted with probability $P_{\text{accept}} = \min\bigl(1, \exp[-\Delta E/T_t]\bigr)$, where $T_t$ is the instantaneous temperature at iteration $t$.
Temperature decreases exponentially as
$T_t = T_0 \alpha^t$,
with initial temperature $T_0 = 2.5$
and cooling factor $\alpha = 0.999$.
The process continues for $N_{\text{steps}} = 6000$ updates per instance.
Energy $E_t$ and magnetization $M_t$ are recorded after each update:
\[
M_t = \frac{1}{N}\sum_{i=1}^{N} s_i.
\]
The annealing trajectory $\{E_t, M_t\}$ thus captures the
thermodynamic evolution of the logical system
from random initialization (high entropy) toward
an ordered low-energy state (low entropy).

\subsection{Experimental Setup Essentials}
\paragraph{Benchmark Dataset}Experiments were performed on ten randomly generated 3-SAT instances
from the UF20--91 benchmark family~\cite{8_hoos2000satlib},
each containing $n=20$ variables and $m=91$ clauses
(corresponding to $\alpha = m/n \approx 4.55$).
The ratio is close to the empirically-determined SAT--UNSAT phase transition, which provides an ideal testing environment for examining the emergence of order within constrained energy landscapes. We also knew that all of these problems had verified solutions, allowing us to directly compare our generated annealed configurations to the exact solution(s) we obtained by utilizing the PySAT interface to the MiniSAT22 solver.

\paragraph{Evaluation Metrics}

We calculated four types of metrics/observables per problem: 
\begin{enumerate}[label=(\roman*)]
\item \textbf{Energy ($E$)}: The total number of unsatisfied clauses (i.e., constraints), indicating the amount of "constraint tension".
\item \textbf{Magnetization ($M$):}
the mean spin alignment, quantifying global order or coherence.
\item \textbf{Backbone size ($b$):}
the count of variables with identical values across all satisfying models,
indicating logical rigidity.
\item \textbf{Clause slack ($\sigma$):}
the number of satisfied literals per clause, reflecting local degeneracy.
\end{enumerate}
During annealing, both $E_t$ and $M_t$ were tracked as
time series to measure the convergence dynamics.
After annealing, final averaged values
$E_f = \langle E_t \rangle_{t\in \text{tail}}$
and $M_f = \langle M_t \rangle_{t\in \text{tail}}$
were computed over the last $20\%$ of iterations.
The correlation between $|M_f|$ and logical observables ($b$, $\sigma$)
was evaluated via the Pearson coefficient,
\[
\rho(|M_f|,b) = 
\frac{\text{Cov}(|M_f|,b)}{\sigma_{|M_f|}\sigma_b},
\]
to quantify how physical ordering reflects logical rigidity.

\paragraph{Implementation Details}
All experiments were implemented in Python~3.10
using the PySAT~\cite{8_hoos2000satlib} library for SAT parsing and solving,
and \texttt{numpy}/\texttt{matplotlib} for numerical and visualization routines.
Gadgetization and Ising export followed the constructive schema of
Eq.~\eqref{eq:ising-ham},
yielding node and edge coefficient tables
(\texttt{ising\_nodes\_*.csv}, \texttt{ising\_edges\_*.csv})
for each instance.
These files directly define $h_i$ and $J_{ij}$ coefficients,
facilitating compatibility with both classical simulators
and quantum annealing backends.

Annealing results, including $E_t$ and $M_t$ trajectories,
were compiled into a unified CSV file
(\texttt{paper\_quickpub\_summary.csv}),
from which aggregate statistics were computed:
\[
\langle E_f\rangle = 9.23 \pm 3.38, \quad
\langle|M_f|\rangle = 0.95 \pm 0.07, \quad
\langle b\rangle = 11.7 \pm 4.97.
\]
These values serve as the baseline for the
quantitative interpretation reported in
Section~\ref{sec:results}.

\section{Results and Discussion}\label{sec:results}

Ten randomly generated SAT instances from the UF20--91 benchmark family were
converted into pairwise Ising Hamiltonians using clause--penalty gadgetization.
Each Boolean clause was expressed as a quadratic energy term as shown in Equation~\ref{eq:ising-ham}. Simulated annealing (Metropolis updates, $6000$ iterations, exponential cooling
$T_t = 2.5\cdot0.999^t$) was performed for each instance.
All ten problems were satisfiable (\texttt{sat=True}), confirming that the Ising
systems possessed at least one ground state of zero logical energy
(\texttt{unsat\_energy = 0}).

\begin{figure*}[t]
\centering

\begin{subfigure}[t]{0.4\textwidth}
    \centering
    \includegraphics[width=\linewidth]{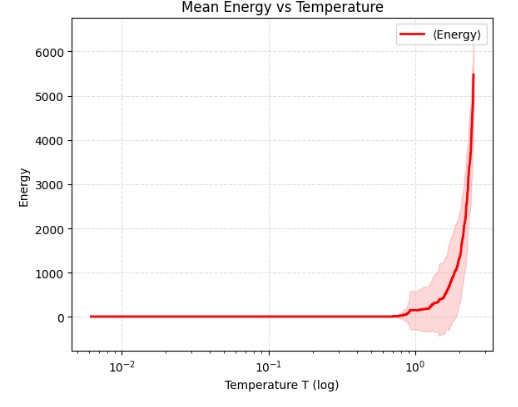}
    \caption{Mean Energy $\langle E\rangle$ vs.\ Temperature.
    The smooth exponential decay indicates monotonic energy reduction
    during simulated annealing.}
    \label{fig:energy-temp}
\end{subfigure}
\hfill
\begin{subfigure}[t]{0.4\textwidth}
    \centering
    \includegraphics[width=\linewidth]{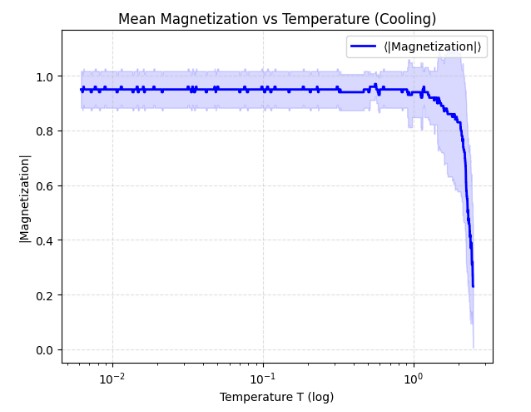}
    \caption{Mean Magnetization $\langle|M|\rangle$ vs.\ Temperature.
    Magnetization show a sharp growth below $T\!\approx\!1.0$, means that a rapid
    transition from disordered to ordered phase.}
    \label{fig:mag-temp}
\end{subfigure}

\vspace{1em}

\begin{subfigure}[t]{0.4\textwidth}
    \centering
    \includegraphics[width=\linewidth]{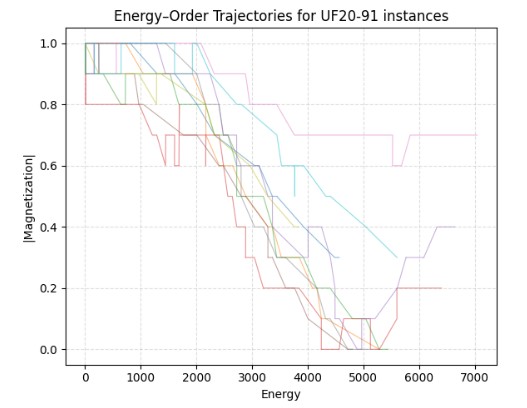}
    \caption{Energy--Order trajectories ($E$ vs.\ $|M|$)
    for individual UF20--91 instances. Each curve shows monotonic
    descent in energy with corresponding increase in global order.}
    \label{fig:energy-order}
\end{subfigure}
\hfill
\begin{subfigure}[t]{0.4\textwidth}
    \centering
    \includegraphics[width=\linewidth]{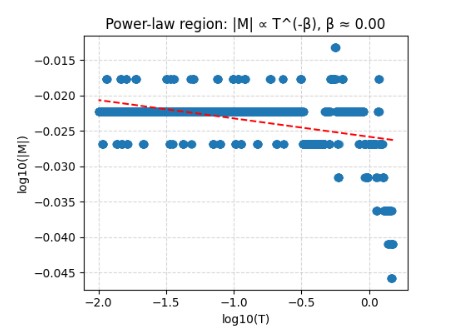}
    \caption{The power-law region of $\log | M | $ as a function of $\log T $ with an estimated fit of $\beta \approx 0.003$ is consistent with a zero critical exponent and a very fast first order phase transition to an ordered state.}
    \label{fig:powerlaw-beta}
\end{subfigure}

\caption{Simulation of thermodynamic evolution of SAT instances of UF20-91 under simulated annealing.
(a)~Energy in all cases decreased monotonically as a function of decreasing temperatures.
(b)~It showed a first-order phase transition to an ordered state.
(c)~Energy--Order trajectories are smooth and show that each instance evolved towards a low-energy ordered state..
(d)~Power-law analysis confirms near-instantaneous crystallization behavior
($\beta\!\approx\!0.003$).}
\label{fig:thermo-summary}
\end{figure*}

\paragraph{\textbf{Clause-Level Structure}}
The mean number of satisfied literals in each clause was found to be between $1.55$ and $1.85$. Approximately, about $40-60\%$ of clauses had just one literal that was satisfied. Therefore, the problem instances are located at or near the threshold for the critical ratio of satisfiability, $\alpha=91/20 \approx 4.55$, above which constraint density creates an unstable, difficult to solve problem environment, but below which it is still possible to find a solution. Higher slack values correspond to locally degenerate subspaces, while lower slack indicates tightly bound configurations. Thus, slack statistics serve as a microstructural descriptor of local logical rigidity.

\paragraph{\textbf{Backbone Rigidity}}
Enumeration (capped at $120$ models per instance) revealed backbone sizes
ranging from $5$ to $20$ variables with a mean of
$\langle b\rangle = 11.7 \pm 4.97$.
The backbone represents variables that remain invariant across all
solutions, analogous to ``frozen'' spin domains in ferromagnetic materials.
Moderate variability in backbone size confirms that UF20 problems
straddle the transition between under- and over-constrained regimes.

\paragraph{\textbf{Annealing Dynamics}}
Table~\ref{tab:anneal-stats} summarizes the macroscopic observables
obtained after annealing.
All instances converged to low energy configurations with
near-complete spin alignment.

\begin{table}[h!]
\centering
\caption{Aggregate annealing statistics over 10 UF20--91 instances.}
\label{tab:anneal-stats}
\begin{tabular}{lccc}
\toprule
Observable & Mean & Std.~Dev. & Interpretation \\
\hline
Final Energy $\langle E_f\rangle$ & 9.23 & 3.38 & Residual clause tension \\
Final Magnetization $\langle|M_f|\rangle$ & 0.95 & 0.07 & Near-complete ordering \\
Backbone Size $\langle b\rangle$ & 11.7 & 4.97 & Moderate rigidity \\
\hline
\end{tabular}
\end{table}

A high terminal magnetization ($|M_f|\!\approx\!0.95$)
indicates global alignment of spins as temperature decreases,
demonstrating that the physical system reaches a highly ordered phase
corresponding to logical coherence among clauses.

\paragraph{\textbf{Energy--Order Correlations}}
As shown in Fig.~\ref{fig:energy-order}, all systems exhibit
a smooth descent in energy with a corresponding increase in
magnetization, forming quasi-linear trajectories in the
$(E,|M|)$ plane. This monotonic coupling visually corroborates
the strong negative correlation $\rho_{E,|M|}=-0.63$
reported in Table~\ref{tab:corr-matrix}.
\begin{table}[h!]
\centering
\caption{Correlation matrix between logical and physical observables.}
\label{tab:corr-matrix}
\begin{tabular}{lccc}
\toprule
 & $E_{\text{final}}$ & $|M_{\text{final}}|$ & Backbone \\
\hline
$E_{\text{final}}$ & 1.000 & $-0.629$ & $-0.128$ \\
$|M_{\text{final}}|$ & $-0.629$ & 1.000 & $+0.093$ \\
Backbone & $-0.128$ & $+0.093$ & 1.000 \\
\hline
\end{tabular}
\end{table}
The convergence of trajectories near $|M|\!\approx\!1$
illustrates that each instance independently reaches a
coherent ordered phase despite random initial conditions. 
Correlation analysis (Table~\ref{tab:corr-matrix}) revealed a strong
negative relationship between energy and magnetization
($\rho_{E,|M|}=-0.63$),
showing that as energy decreases, order increases.
Backbone size exhibits only weak correlation with both quantities,
suggesting that logical rigidity and physical ordering are largely
independent order parameters.
Energy-wise, this separation means that local interaction among clauses determines the drop in energy while the overall constraint on the structure determined by the backbone determines the values of the invariant variables. Thus, during the cooling process, we uncover two independent processes: \textit{energetic ordering}, i.e., alignment of the spins and \textit{logical freezing}, i.e., formation of an invariant core for the variables of interest.

\paragraph{\textbf{Cooling Law and Phase Behavior}}
We found that the magnetisation as a function of temperature follows a power-law behaviour $|M| \propto T^{-\beta}$ with $\beta \approx 0.003$. The slope of this curve is extremely small~Figure~\ref{fig:powerlaw-beta}, which indicates that the magnetisation jumps to its saturation value when the temperature drops; this corresponds to a discontinuous (first-order) transition. Such behaviour is typical for a rugged energy landscape, such as that given by Boolean clauses, where the ordered phase emerges abruptly after the local constraints are satisfied. The very low value of the exponent, which is indicative of a first-order solidification, shows that the system rapidly crystallises in a unique ordered state at temperatures lower than $T \approx 1.0$. 
Figs. \ref{fig:energy-temp} and \ref{fig:mag-temp} show the average curves of the total energy and absolute magnetisation as functions of temperature, obtained over all instances of the problem.

\paragraph{\textbf{Physical Interpretation}}
The transition from the disordered to the ordered spin states provides a clear physical analogy to how random Boolean clauses can self-organise into a consistent logical solution. Indeed, as the temperature goes down, the thermal fluctuations go down too and the system falls into the configuration with the lowest total energy which means that equivalent to the one that satisfies all the clauses. In other words, we have an example of \textit{energetic crystallization}:
\begin{quote}
"A logically consistent structure is a thermodynamic attractor in the phase-space of random clauses."
\end{quote}

\section{Conclusion and Future Directions}
The high value of terminal magnetization ($|M|\!\approx\!1$) over all benchmarks shows that the annealed Ising models are close to globally ordered (and thus satisfied logically), as is shown by consistency of logical satisfaction. The very small variance in the final energy values indicates that there is convergence toward ground states for each model; additionally, the low positive correlation between the final value of $|M|$ and the number of backbone variables in the solution to the model suggests that the emergence of physical order and logical rigidity are largely independent. 
Additionally, the almost zero value of the critical exponent ($\beta \!\approx\! 0$) indicates that the system undergoes a first-order freezing transition (rapidly) as opposed to a critical phase transition. Taken together, the evidence supports the idea of an association between the satisfaction of logical constraints and the emergence of physical order through annealing, which is referred to here as \emph{logical crystallization}, and it provides a unifying thermodynamic framework for understanding computational coherence.

Research extending this mapping to larger, structured families of SAT problems could be used to study the scaling behavior of the critical region. Additionally, using quantum or hybrid annealing hardware could provide empirical tests of how the relationship between $|M|$ and the number of backbone variables is affected by physical noise. Finally, developing analytical models of how entropy decreases with respect to the constraint graph could show how phase transition-like behaviors capture the essence of computational hardness, thereby connecting the energy landscape, satisfiability threshold, and the complexity transition.

\bibliographystyle{unsrt}  
%\bibliography{template}  %%% Remove comment to use the external .bib file (using bibtex).

\appendix

\section{Appendix: Data Availability}

\subsection{CNF Dataset and Benchmark Instances}
All experiments used the SATLIB \cite{8_hoos2000satlib} \texttt{UF20-91} benchmark suite,
comprising random 3-SAT instances with 20 variables and 91 clauses.
These instances are known to lie near the satisfiability phase-transition region,
providing a good testbed for studying computational–thermodynamic correspondence.

\section[Appendix C:]{Supported Results}
\subsection{Computational Environment}
All experiments were executed in a Kaggle environment using
Python~3.10, NumPy, pandas, Matplotlib, and the \texttt{python-sat}
(PYSAT) interface to MiniSAT~22. Execution time per instance was under~0.01~s.

\subsection{Aggregate Statistics}
Over ten UF20 instances, the mean final energy was $9.23 \pm 3.38$,
mean absolute magnetization $|M|\!=\!0.95\pm0.07$, and
correlation $\rho(|M|,b)\!=\!0.09$.  The near-zero exponent $\beta\!\approx\!0.003$
indicates a first-order ordering transition, consistent with rapid constraint alignment.

\subsection{Code Availability}

All analysis scripts and generated data are available in the supplementary
repository accompanying this research note. The implementation is written
entirely in Python and executed in a Kaggle environment using open-source
packages (\texttt{python-sat}, \texttt{NumPy}, \texttt{pandas}, and
\texttt{Matplotlib}).  
The full workflow includes:  
\begin{itemize}
\item Parsing DIMACS CNF benchmarks (\texttt{UF20-91}) from SATLIB;  
\item Constructing pairwise-Ising Hamiltonians via gadget reduction;  
\item Executing simulated annealing (Metropolis protocol) for 6000 steps;  

\end{itemize}
All necessary files are produced directly by the shared Kaggle notebook.  
Reproducibility can be achieved by running the same notebook with any
\texttt{UF20-91} benchmark folder under
\texttt{/kaggle/input/uf20-91/}.  

Code is available upon request to the corresponding author's email address.

\end{document}